\documentclass[fleqn,usenatbib]{mnras}

\usepackage{newtxtext,newtxmath}

\usepackage[T1]{fontenc}
\usepackage{tablefootnote}

\DeclareRobustCommand{\VAN}[3]{#2}
\let\VANthebibliography\thebibliography
\def\thebibliography{\DeclareRobustCommand{\VAN}[3]{##3}\VANthebibliography}

\usepackage{graphicx}	
\usepackage{amsmath}
\usepackage{ulem}

\newcommand{\fermi}{{\it Fermi}-LAT}
\newcommand{\gray}{$\gamma$-ray}
\newcommand{\grays}{$\gamma$-rays}
\newcommand{\source}{CTA 102}
\newcommand{\gsim}{{\lower.5ex\hbox{$\; \buildrel > \over \sim \;$}}}

\title[Broadband emission from CTA 102]{Modeling the time variable spectral energy distribution of the blazar CTA 102 from 2008 to 2022}

\author[N. Sahakyan et al.]{
N. Sahakyan$^{1,2, 3}$, \thanks{E-mail: narek@icra.it}
D. Israyelyan$^{1}$,
G. Harutyunyan$^{1, 4}$,
S. Gasparyan$^{1}$,
V. Vardanyan$^{1}$,
M. Khachatryan$^{1}$\\
$^{1}$ICRANet-Armenia, Marshall Baghramian Avenue 24a, Yerevan 0019, Armenia\\
$^{2}$ICRANet, P.zza della Repubblica 10, 65122 Pescara, Italy\\
$^{3}$ ICRA, Dipartimento di Fisica, Sapienza Universita` di Roma, P.le Aldo Moro 5, 00185 Rome, Italy\\
$^{4}$ Byurakan Astrophysical Observatory, Aragatsotn reg., Armenia
}

\date{Accepted XXX. Received YYY; in original form ZZZ}
\pubyear{2022}

\begin{document}
\label{firstpage}
\pagerange{\pageref{firstpage}--\pageref{lastpage}}
\maketitle

\begin{abstract}
We present long-term multiwavelength observations of blazar CTA 102 ($z=1.037$). Detailed temporal and spectral analyses of $\gamma$-ray, X-ray and UV/optical data observed by {\it Fermi}-LAT, Swift XRT, NuSTAR and Swift-UVOT over a period of 14 years, between August 2008 and March 2022, was performed. We found strong variability of source emission in all the considered bands, especially in the $\gamma$-ray band it exhibited extreme outbursts when the flux crossed the level of $10^{-5}\:{\rm photon\:cm^{-2}\:s^{-1}}$. Using the Bayesian Blocks algorithm, we split the adaptively binned $\gamma$-ray light curve into 347 intervals of quiescent and flaring episodes and for each period built corresponding multiwavelength spectral energy distributions (SEDs), using the available data. Among the considered SEDs, 117 high-quality (quasi) contemporaneous SEDs which have sufficient multiwavelength data, were modeled using JetSeT framework within a one-zone leptonic synchrotron and inverse Compton emission scenario assuming the emitting region is within the broad-line-region and considering internal and external seed photons for the inverse Compton up-scattering. As a result of modeling, the characteristics of the relativistic electron distribution in the jet as well as jet properties are retrieved and their variation in time is investigated. The applied model can adequately explain  the assembled SEDs and the modelling shows that the data in the bright flaring periods can be reproduced for high Doppler boosting and magnetic field. The obtained results are discussed in the context of particle cooling in the emitting region.
\end{abstract}

\section{Introduction}\label{intro}
Active galactic nuclei powered by supermassive black holes with masses of $10^6$-$10^{10}\:M_{\odot}$ are the most luminous persistent objects in the extragalactic sky. In some AGNs a relativistic jet is formed perpendicular to the accretion disc plane and it plays a crucial role in blazar classifications. According to the unification scheme developed by \citet{1995PASP..107..803U}, an AGN is called a blazar when the jet is closely aligned with the line of sight of the observer. Blazars are characterized by high radio and optical polarization, apparent superluminal motion along with high-amplitude variability in all accessible bands of the electromagnetic spectrum. Usually, this variability is unpredictable and only for a few objects periodic variability is observed \citep[e.g., see][]{2015ApJ...813L..41A, 2022arXiv220413051R}. Blazars are usually believed to be persistent sources, however recently a blazar showing a transient behaviour was observed. Namely, 4FGL J1544.3-0649 was never detected in the X-ray and \gray\ bands until May 2017 when it rose above the detectability level and for a few months became one of the brightest X-ray blazars \citep{2021MNRAS.502..836S}. Blazar emission is dominated by non-thermal emission from the jet which is significantly Doppler amplified since the jet with superluminal motion is viewed at small angles. Because of this, blazars even at higher redshift are observed \citep[e.g., see][]{2020MNRAS.498.2594S}.

The emission from blazars is observed in a wide frequency range, from radio to high energy (HE; $>100$ MeV) and very high energy (VHE; $>100$ GeV) \gray\ bands \citep{2017A&ARv..25....2P} displaying a double hump structure in their broadband spectral energy distribution (SED). The first component (low-energy) usually peaks between far infrared and X-rays while the second component (HE) is observed between X-rays and VHE \grays. The low-energy component is explained by the synchrotron emission of jet-accelerated electrons under the magnetic field while the origin of the HE component is discussed within leptonic and hadronic models, depending on the type of emission initiating particles, $e^{-}e^{+}$ pairs or hadrons. According to the widely discussed leptonic scenario, the HE component is due to inverse Compton upscattering of photons by energetic electrons. Most common scenarios used in the literature are synchrotron self-Compton (SSC) model and the external Compton (EIC) model. According to the first scenario, the internal synchrotron photons are up-scattered to higher energies \citep{ghisellini, bloom, maraschi} whereas the latter model assumes the photons are produced external to the jet \citep{blazejowski,ghiselini09, sikora}. In alternative hadronic or lepto-hadronic scenarios, the protons co-accelerated with the electrons make a non-negligible contribution to the HE component. This contribution can be either directly from proton synchrotron radiation \citep{2001APh....15..121M} or from secondaries produced in the proton-photon interactions or photo-pair productions \citep{1993A&A...269...67M, 1989A&A...221..211M, 2001APh....15..121M, mucke2, 2013ApJ...768...54B, 2015MNRAS.447...36P, 2022MNRAS.509.2102G}. Lately, the hadronic models \citep[especially lepto-hadronic][]{2022MNRAS.509.2102G} have become more attractive after the detection of VHE neutrinos spatially coinciding with the direction of known blazars \citep{2018Sci...361..147I, 2018Sci...361.1378I, 2018MNRAS.480..192P}. The initial association between TXS 0506+056 and IceCube-170922A event provided first multimessenger picture of blazar and opened a wider perspective for theoretical studies \citep[][]{2018ApJ...863L..10A,2018ApJ...864...84K, 2018ApJ...865..124M, 2018MNRAS.480..192P, 2018ApJ...866..109S, 2019MNRAS.484.2067R,2019MNRAS.483L..12C, 2019A&A...622A.144S, 2019NatAs...3...88G, 2022MNRAS.509.2102G}. The assumption that blazars are neutrino sources was further strengthened by the observation of multiple neutrino events from the direction of PKS 0735+178 when the source was undergoing a major flaring activity in the optical/UV, X-ray and \gray\ bands \citep{2022arXiv220405060S}.\\
\indent Commonly, the blazars are grouped based on their optical spectral properties. Namely, Flat Spectrum Radio Quasars (FSRQs) which show strong optical lines and BL Lacertae type objects (BL Lacs) which have very faint optical emission lines. Blazars are further classified based on the observed SEDs. Namely, based on the frequency where the synchrotron component peaks ($\nu_{\rm p}$), blazars are separated into low, intermediate and high-energy peaked sources \citep{Padovani1995,Abdo_2010}; low synchrotron peaked sources (LSPs or LBLs) when $\nu_{\rm p}<10^{14}$ Hz, intermediate synchrotron peaked sources (ISPs or IBLs) when $10^{14}\:{\rm Hz}\:<\nu_{\rm p}<10^{15}$ Hz and high synchrotron peaked sources (HSPs or HBLs) when $\nu_{\rm p}>10^{15}$ Hz. 
However, recently \citet{2021Univ....7..492G} showed that there are strong similarities between the properties of IBLs and HBLs and they show large differences from LBLs, so the classification can be refined into LBLs and  intermediate-high-energy-peaked objects (IHBLs) when $\nu_{\rm p}$ is below or above $10^{13.5}$ Hz.\\
\indent \source\ is a FSRQ with a redshift of $z = 1.037$ \citep{1965ApJ...141.1295S}. Harboring a black hole with a mass of $8.5\times10^8\:{\rm M_{\rm BH}}$ \citep{2014Natur.510..126Z}, \source\ is one of the brightest FSRQs observed in the HE \gray\ band. It was initially observed by the Compton Gamma Ray Observatory mission having estimated a \gray\ flux of $(2.4\pm0.5)\times10^{-7}\:{\rm photon\:cm^{-2}\:s^{-1}}$ \citep{1993ApJ...414...82N}. Then, \source\ was scanned continuously by the Fermi Large Area Telescope (\fermi) since mid-2008, initially showing that the source is relatively weak in the \gray\ band. However, from 2016 to 2017 it underwent an unprecedented outburst in all the wavebands \citep{casadio, Balonek, chapman, popov, ciprini, bulgarelli, ciprini1, becerra, minervini, carrasco}. For example, in the \gray\ band its flux was as high as $(3.55\pm0.55)\times10^{-5}\:{\rm photon\:cm^{-2}\:s^{-1}}$ \citep{2018ApJ...863..114G} and in some active \gray\ periods its spectrum also deviated from simple power-law model \citep{2020A&A...635A..25S}. During the \gray\ flares, the source was so bright that variability was investigated down to minute scales \citep{2018ApJ...854L..26S}. In December 2016, the source was also in an extreme optical and near-IR out-bursting state when the brightness increased up to six magnitudes with respect to the faint state of the source \citep{2017Natur.552..374R}. Various theoretical models were used to explain the flaring behaviour of \source\ which includes an inhomogeneous curved jet with different jet regions changing their orientation and consequently the Doppler factors \citep{2017Natur.552..374R}, or a superluminal component crossing a recollimation shock \citep{2019A&A...622A.158C}, or lepto-hadronic processes when the gas cloud penetrates the jet \citep{2017ApJ...851...72Z, 2019ApJ...871...19Z} or the activities were interpreted as change of the location of the emission region \citep[e.g.,][]{2018ApJ...863..114G, 2018ApJ...866...16P, 2020A&A...635A..25S}, etc.

Due to the long-lasting and peculiar multiwavelength flaring activity, \source\ was frequently observed in different bands and became one of the most-studied blazars \citep[][]{2018ApJ...854...17L, 2018A&A...617A..59K, 2019MNRAS.490.5300D, 2019ApJ...877...39M, 2020ApJ...891...68C, 2021MNRAS.500.5297A, 2021MNRAS.504.1103R, 2022MNRAS.510..815K, 2022ApJS..260...48G}. Although many studies have been conducted which lead to a better understanding of the \source\ jet, it is up to now not clear the origin of the multiwavelength flares of \source, especially the changes in the jet that have led to prolonged flaring activities.

The monitoring of \source\ during its unprecedented outburst with various instruments resulted in accumulation of an extensive data set. In addition, before and after the outburst the source was also monitored in the \gray\ band with \fermi\ and observed in the optical/UV and X-ray bands by Neil Gehrels Swift Observatory \citep{2004ApJ...611.1005G}, (hereafter Swift). This can be combined with other available data to build the broadband SEDs of \source\ in various (flaring or quiescent) periods with (quasi) contemporaneous data. These SEDs with various spectral properties represent an ample variety of source emission in different states and their modeling is crucial for understanding of the physical processes and their changes in time. In the broadband SEDs of blazars the changes are expected to be due to the variation of the parameters of the emitting electrons or the physical parameters of the emission region. Therefore, the modeling of the SEDs in different periods allows to connect the observational properties with the physical processes at work in jets. For example in \citet{2021MNRAS.504.5074S} and \citet{2022MNRAS.513.4645S} the modeling of a large number of contemporaneous SEDs of 3C 454.3 and BL Lac allowed to estimate the main parameters describing the emitting electrons and the emission region and investigate their evolution in time which was crucial for understanding of the observed spectral changes in them.

Motivated by the availability of multiwavelength data from \source\ observations  before, during and after the large outburst, for furthering our knowledge of the emission processes dominating in the jet of \source\ we performed an intense broadband study of \source\ using the data accumulated during 2008-2022. We have systematically investigated the spectral and variability properties of the source emission in the optical/UV, X-ray and \gray\ bands. We performed a deep investigation of the origin of the source emission in various periods by generating as many SEDs of \source\ as possible that can be constructed with contemporaneous data and modeling them within the leptonic scenario. The paper is structured as follows. The broadband data analyses are described in Section \ref{mwl}. The multiwavelength variability is explored in Section \ref{mw_var}. The modeling of broadband SED is described in Section \ref{mw_sed}. We present the discussion and results in Section \ref{res} and the conclusions in Section \ref{conc}.
\section{Multiwavelength observations of \source}\label{mwl}
Exhibiting interesting multiwavelength properties, \source\ was frequently observed in different bands. Below we report the data analyzed in this paper or extracted from public archives which was used in the current study.

\subsection{\fermi\ observations of \source}
Operating since 2008, \fermi\ provides an exceptional view of the \gray\ sky, imaging the entire sky every three hours \citep{2009ApJ...697.1071A}. In the current paper the \fermi\ data accumulated between 04 August 2008 and 04 March 2022 in the 100 MeV–300 GeV range were downloaded and analyzed using the Fermi ScienceTools  version 2.0.8 and P8R3\_SOURCE\_V3 instrument response function.
Events were extracted from a region of interest (ROI) with a $12^\circ$ radius centered on the source position (RA: 338.15, DEC: 11.73). As recommended by the \fermi\ team, the cut {\it evclass = 128} and {\it evtype=3} was applied to select events with higher probability of being photons. Whereas, the filter ${\rm (DATA\_ QUAL>0)\&\& (LAT\_ CONFIG==1)}$ was applied to update the good time interval based on spacecraft specifications. A maximum zenith angle cut of $> 90^\circ$ is applied to reduce the contamination from Earth limb \grays. The model file was generated based on \fermi\ fourth source catalog Data Release 3 \citep[4FGL-DR3; ][]{2020ApJ...892..105A, 2022ApJS..260...53A} which includes point sources within the ROI and standard Galactic (gll\_ iem\_ v07) and the isotropic (iso\_P8R3\_SOURCE\_ V3\_v1) diffuse emission components. The spectral parameters of the background sources falling between $12^{\circ}$ and $12^{\circ}$+$5^{\circ}$ were fixed to the values published in the 4FGL-DR3 catalog, while the parameters of the other sources (within $12^{\circ}$) and background models were left free. The best match between the source parameters and the data was obtained by applying standard binned likelihood analysis with {\it gtlike} tool.

After analyzing the data accumulated in the whole time interval, light curves were computed with different time bins to investigate the variability in the \gray\ band. Initially, the entire period was divided into three-day intervals (1653 in total) and for each single period the flux, photon index (\source\ spectrum was modeled with power-law distribution) and the Test Statistics \citep[TS, defined as twice the difference between the log-likelihoods of the model computed with and without including the source;][]{1996ApJ...461..396M} were estimated.

Next, for a deeper investigation of the \gray\ flux variability, the light curve was generated with the help of the adaptive binning method \citep{2012A&A...544A...6L}. As distinct from the fixed time interval light curve where the longer bins will smooth out the fast variation and in short time intervals the flux can be estimated only in the bright state of the source, in the adaptively binned light curve the bin width is defined by requiring a constant relative flux uncertainty above an optimal energy, so the time bins are longer during low flux levels and narrower when the source is in flaring state. This allows to track the evolution of the \gray\ flux in time, extract maximum possible information and identify flaring periods \citep[e.g., see][]{2018ApJ...863..114G, 2018A&A...614A...6S, 2017MNRAS.470.2861S, 2017A&A...608A..37Z, 2017ApJ...848..111B, 2016ApJ...830..162B, 2013A&A...557A..71R}.

The spectral changes in the \gray\ band were further investigated by producing the source spectrum in different periods. For this purpose, the adaptively binned light curve is divided into piece-wise constant blocks \citep[Bayesian blocks][] {2013ApJ...764..167S} representing optimal segmentation of the data into time intervals during which the flux is constant. By this approach, the considered period is divided into 347 intervals with the same flux level, whether flaring or quiescent. The spectrum of \source\ in each of the selected period is computed by applying unbinned likelihood analysis and running {\it gtlike} separately for 5 (when the source is in average or quiescent state) or 7 energy bins (when the source is in flaring state) of equal width in log scale.

\begin{figure*}
	\includegraphics[width=0.9\textwidth]{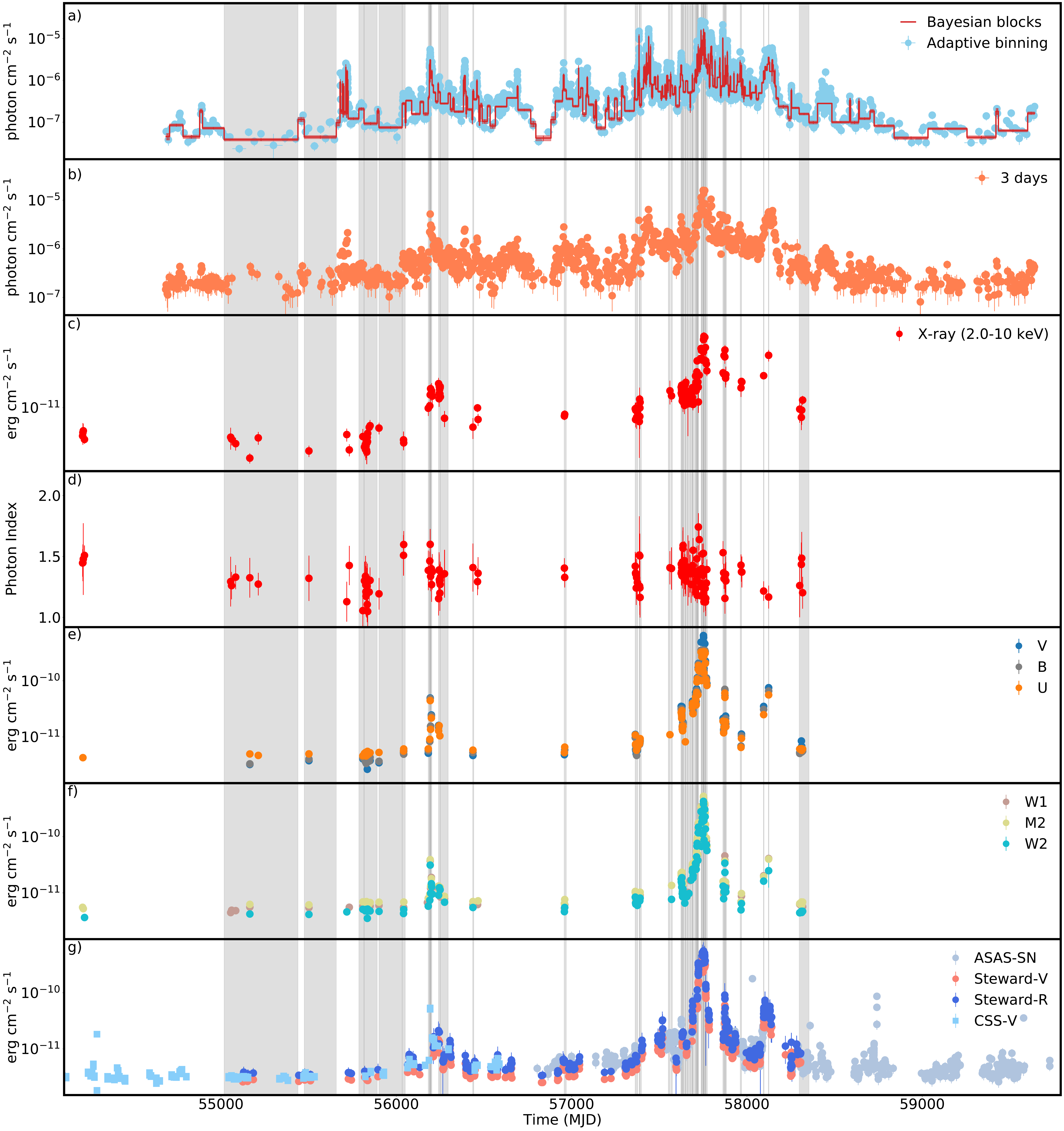}
    \caption{The multiwavelength light curve of \source\ between 2007 and 2022. \textit{a)} Adaptively binned \gray\ light curve ($>166.3$ MeV) with the Bayesian blocks, \textit{b)} 3-day binned \gray\ light curve, \textit{c)} 2.0-10 keV X-ray flux, \textit{d)} 0.3-10.0 keV X-ray photon index, \textit{e)} flux in V, B, and U filters, \textit{f)} flux in  W1, M2 and W2 filters and \textit{g)} V-band and R-band fluxes. The periods for which the SEDs have been modeled are highlighted in gray.}
    \label{lightcurve_all}
\end{figure*}
\subsection{Swift observations of \source}
In the optical/UV and X-ray bands there are available a total of 146 observations of \source\ with Swift XRT/UVOT instruments. All the XRT observations were individually downloaded and analyzed using {\it Swift\_xrtproc} pipeline \citep{2021MNRAS.507.5690G}. This tool developed within the Open Universe Initiative downloads the raw data and calibration files from one of the official Swift archives, processes it using the XRTPIPELINE task for each snapshot and for the entire Swift observation, applies pile-up correction when the source count rate is above $0.5\:{\rm counts\: s^{-1}}$ and generates source (from a circle with a radius of 20 pixels centered at the position of the source) and background (an annular ring centered at the source) spectral files. It performs a spectral fitting with XSPEC (version 12.12.0) on the ungrouped data using Cash statistics \citep{1979ApJ...228..939C}, modeling \source\ spectrum as a power-law and a log-parabola. As a result, the tool generates SED data and estimates the flux and photon index in various bands. More details on {\it Swift\_xrtproc} are given in \citet{2021MNRAS.507.5690G}.

The Swift-UVOT data in three optical filters (V, B, and U) and three UV filters (W1, M2, and W2) were downloaded and reduced using HEAsoft version 6.29 with the latest release of HEASARC CALDB. The source counts were extracted from a region of 5 arcsec radius centered at the source and the background counts from a region of 20 arcsec centered away from the source. {\it uvotsource} tool was used to obtain the magnitude which was corrected for reddening and galactic extinction using the reddening coefficient $E(B - V)$ from the Infrared Science Archive \footnote{http://irsa.ipac.caltech.edu/applications/DUST/}.
\subsection{NuSTAR observations of \source}
NuSTAR with two focal plane modules \citep{2013ApJ...770..103H}, FPMA and FPMB, observed \source\ in the hard X-ray (3-79 keV) band on December 30, 2016 for a total exposure of 26.2 ksec. The NuStar data was processed with {\it NuSTAR\_ Spectra} script which is a shell script based on the \textit{NuSTAR} Data Analysis Software (NuSTARDAS) that automatically downloads calibrated and filtered event files from the SSDC repository, generates scientific products and carries out a complete spectral analysis. It uses \textit{nuproducts} to generate the spectra from source counts extracted from a circular region whose radius is set to a value that is optimised depending on the source count rate ($30''$ in this case), while the background counts are from an annulus centered on the source. With the XSPEC, the spectral analysis is performed adopting Cash statistics for the energy range from 3 keV up to the maximum energy where the signal is still present, typically between 20 and 79 keV. {\it NuSTAR\_ Spectra} script is presented and described in \citet{2022MNRAS.514.3179M}. 
\subsection{Archival optical data}
In order to monitor the flux changes in the optical band, the light curves from several public archives were used. Namely, the optical data (V- and R- band) from Steward Observatory \citep{2009arXiv0912.3621S}, V-band data from the All-Sky Automated Survey for Supernovae (ASAS-SN) \footnote{https://asas-sn.osu.edu/} \citep{2017PASP..129j4502K} and the V-band data from the Catalina Sky Survey \citep[CSS;][]{2009ApJ...696..870D} were downloaded from the public archives.
\section{Multiwavelength variability}\label{mw_var}
The multiwavelength light curve of \source\ is shown in Fig. \ref{lightcurve_all}. The adaptively binned light curve above $166.3$ MeV in Fig. \ref{lightcurve_all} panel a) shows the continuous observation of the source in the \gray\ band and reveals the complex flux changes. During the considered periods several outbursts are observed. Until April 2011 the source flux was constant, not exceeding $10^{-7}\:{\rm photon\:cm^{-2}\:s^{-1}}$. The first flare (when $F_{\rm \gamma}>15\times F_{\rm \gamma, min}$) occurred in April-June 2011 (MJD 55680-55730), when the flux increased up to $(2.55\pm0.62)\times10^{-6}\:{\rm photon\:cm^{-2}\:s^{-1}}$. Other enhancements were observed between September-October 2012 (MJD 56180-56230) and between March-April 2013 (MJD 56380-56400). Yet, a major flaring activity, when the source flux increased above $10^{-5}\:{\rm photon\:cm^{-2}\:s^{-1}}$, was observed between December 2015- March 2016 (MJD 57370-57470). Then, the source entered a prolonged out-bursting state between November 2016 -June 2017 (MJD 57710-57910) when the highest flux of $(2.64\pm0.60)\times10^{-5}\:{\rm photon\:cm^{-2}\:s^{-1}}$ above $166.3$ MeV was observed on MJD 57738.5. Another brightening of the source (although with lower amplitude) was observed between November 2017-March 2018 (MJD 58080-58180).  During the considered period, the \gray\ flux of \source\ was above $10^{-5}\:{\rm photon\:cm^{-2}\:s^{-1}}$ for 121.1 hours in total. The ratio between the highest and lowest fluxes is $\simeq1137$ which again shows the high-amplitude variation of the \gray\ flux. The overall trends revealed in the \gray\ light curve generated by the adaptive binning method are also visible in the 3-day light curve (panel b) Fig. \ref{lightcurve_all}) but, as expected, the intra-day flux variability is smoothed out. 

\begin{figure}
	\includegraphics[width=0.48\textwidth]{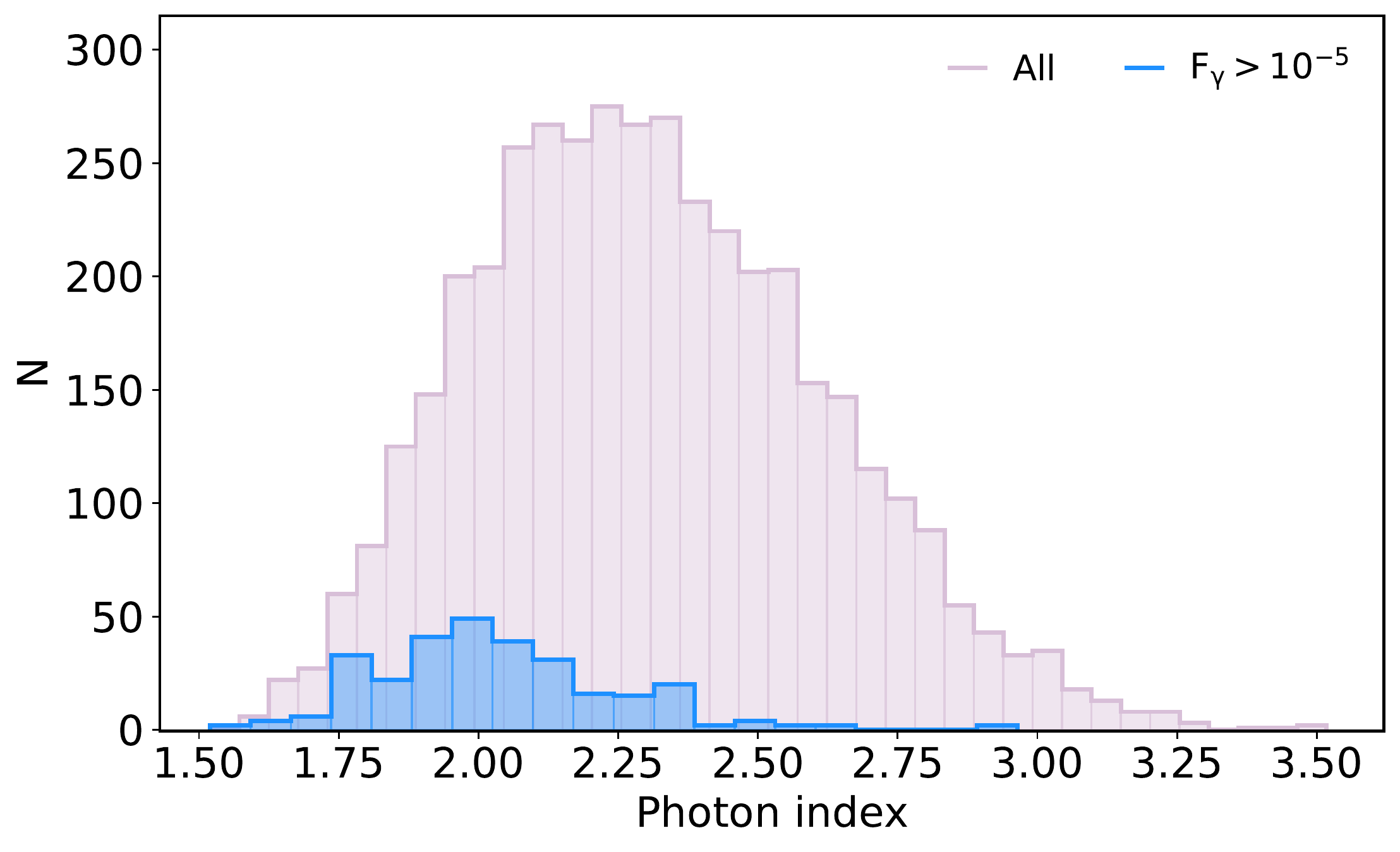}
    \caption{The distribution of the \gray\ photon index estimated in the adaptively binned intervals. The light red area shows the total distribution, while the blue is only when the \gray\ flux was above $10^{-5}\:{\rm photon\:cm^{-2}\:s^{-1}}$.}
    \label{index_histo}
\end{figure}

Together with the \gray\ flux, the photon index varies as well. The hardest photon index is $\Gamma_{\gamma}=1.52\pm0.12$ observed on MJD 57752.5 when the source was in bright \gray\ state with a flux of $(1.02\pm0.20)\times10^{-5}\:{\rm photon\:cm^{-2}\:s^{-1}}$. The distribution of photon index estimated in all adaptively binned intervals is shown in Fig. \ref{index_histo} (light magenta). The mean of the photon index distribution $2.31$ is the same as the time-averaged photon index of the source in 4FGL DR3 ($\sim2.3$). However, there are 353 periods when the photon index was significantly hard ($<1.9$) which means that the peak of the HE component moved to HEs. In Fig. \ref{index_histo}, the blue area corresponds to photon index distribution only when the \gray\ flux was $10^{-5}\:{\rm photon\:cm^{-2}\:s^{-1}}$ which shows that in some of the bright states the photon index of the source was also hard.

The X-ray flux ($2-10$ keV) variation in time is shown in Fig. \ref{lightcurve_all} panel c). There is significant variability of the X-ray flux in different XRT observations. During the prolonged flaring in the \gray\ band, the source was also in an active X-ray emission state, when the X-ray flux reached ${\rm F_{X-ray}[2-10\:keV]}=(5.77\pm0.63)\times10^{-11}\:{\rm erg\:cm^{-2}\:s^{-1}}$. The NuSTAR observation shows that the source flux in the 3-10 keV band is $(4.46\pm0.02)\times10^{-11}\:{\rm erg\:cm^{-2}\:s^{-1}}$ in agreement with the X-ray flux observed by Swift XRT on the same day (${\rm F_{X-ray}[2-10\:keV]}=(5.30\pm0.47)\times10^{-11}\:{\rm erg\:cm^{-2}\:s^{-1}}$). As the X-ray band corresponds to the rising part of the HE component, the flux in the 10-30 keV band increases being $(9.04\pm0.05)\times10^{-11}\:{\rm erg\:cm^{-2}\:s^{-1}}$. Also, the Swift XRT and NuSTAR observations reveal similar photon indexes in the 0.3-10 keV and 3-30 keV bands, $1.25\pm0.08$ and $1.30\pm0.01$, respectively.    

In Fig. \ref{lightcurve_all} panels e), f) and g), the flux variation in the optical/UV band is shown. In the optical band, the source's emission follows the same trend as in the \gray\ and X-ray bands. Namely, Swift UVOT, ASAS-SN, Steward (V and R band) and CSS observations show that the flux was relatively constant up to MJD 56000 and then increased several times around MJD 56200. However, long-lasting flaring activity was observed between MJD 57400-58000 when the flux in the optical band, as observed with all the considered instruments, was above $10^{-11}\:{\rm erg\:cm^{-2}\:s^{-1}}$. The highest flux of $(6.38\pm0.19)\times10^{-10}\:{\rm erg\:cm^{-2}\:s^{-1}}$ was observed in the V-band on MJD 57751.84 by Swift UVOT. The Swift UVOT observations show that between MJD 57718-57768 (November 2016-January 2017), the source was in an extreme bright state in the optical/UV band when the flux was above $10^{-10}\:{\rm erg\:cm^{-2}\:s^{-1}}$. This is in agreement with the results obtained from \source\ monitoring by the Whole Earth Blazar Telescope (WEBT) \citep{2017Natur.552..374R}. The ASAS-SN monitoring of the source shows that another flaring activity was observed on MJD 58741 and then the source's emission in the optical band was on its regular level. 
\section{Broadband SED modeling}\label{mw_sed}
One of the ways for investigation of the underlying physical processes in the jet is through broadband SED modeling. The SEDs constrained with contemporaneous or quasi-contemporaneous data contain valuable information on the emitting particle spectrum and on the condition of the plasma inside the jet. The evolution of the \source\ SEDs in time (SED/light curve animation) is shown here \href{https://youtu.be/jFNkI_psAjo}{\nolinkurl{youtube.com/jFNkI_psAjo}}. These SEDs were generated by plotting the \gray\ spectra for each of the Bayesian blocks shown in Fig. \ref{lightcurve_all} together with the data available in all other energy bands. In a visually effective way, the temporal changes in the \source\ spectra can be seen by going from one to another interval. This animation shows the high-amplitude and spectral changes in different periods, demonstrating dramatic changes of the \source\ during the prolonged out-bursting period.

In FSRQs, such as \source, a one-zone synchrotron/synchrotron-self Compton (SSC) with an external radiation component is expected to produce the broadband emission. The origin of the external photons depends on the location of the emitting region \citep{2009ApJ...704...38S} and photons directly emitted from the disc \citep{1993ApJ...416..458D, 1992A&A...256L..27D}, emitted from the BLR \citep{sikora} or emitted from the dusty torus \citep{blazejowski} can inverse Compton up-scatter and explain the second component in the broadband SED. In the current study we assume that the emitting region is at $10^{17}$ cm distance from the black hole within the BLR and the external photons are the photons emitted from the BLR. The SED modeling when different locations of the emitting region are considered is presented in \citet{2018ApJ...863..114G} and \citet{2020A&A...635A..25S}.

Here, we consider a one-zone leptonic model of jet emission, assuming the accelerated electrons (protons) are injected in the spherical region of radius $R$. This magnetized region with a field strength of $B$ moves along the jet with a bulk Lorentz factor of $\Gamma_{\rm jet}$ at an angle of $\theta$ relative to the observer’s line of sight. As the jet is almost aligned to the observer (small $\theta$), the emission is Doppler boosting by a beaming factor of $\Gamma_{\rm jet}=\delta$. It is assumed that the spectrum of the injected electrons is described by a power-law with an exponential cutoff energy distribution defined as
\begin{align}
    N(\gamma_{\rm e})=N_{0}\:\gamma_{\rm e}^{-p}\:Exp(-\gamma_{\rm e}/\gamma_{\rm cut}),\:\:\:\:\:\:\: \gamma_{\rm e}>\gamma_{\rm min}
\label{eldist}
\end{align}
where $\gamma_{\rm cut}$ and $\gamma_{\rm min}$ are the cut-off and minimum energy of the electrons, respectively, and $p$ is the power-law index of the electron energy distribution. The normalization constant $N_{0}$ defines the energy density of the electrons: $U_{\rm e}=m_{\rm e}c\int\gamma_{\rm e}N(\gamma_{\rm e})d\gamma_{\rm e}$.

In this scenario, the first peak in the SED is described by synchrotron radiation as a consequence of the interaction of relativistic electrons inside the emitting region with the magnetic field. Instead, the second peak (from X-ray to HE \grays) is formed by the contribution of inverse Compton scattering of synchrotron (SSC) and BLR emitted (EIC) photons. The BLR radius and luminosity of \source\ are $R_{\rm BLR}=6.73\times10^{17}\:{\rm cm}$ and $L_{\rm BLR}=4.14\times10^{45}\:{\rm erg\:s^{-1}}$ \citep{2005MNRAS.361..919P}, respectively, and the BLR is modeled as a spherical shell with a lower boundary of $R_{\rm in, BLR}=0.9\times R_{\rm BLR}=6.06\times10^{17}$ cm and an outer boundary of $R_{\rm out, BLR}=1.2\times R_{\rm BLR}=8.08\times10^{17}$ cm. Assuming that the 10\% of the disc luminosity is reprocessed into BLR radiation, the disc luminosity would be $L_{\rm disc}=4.14\times10^{46}\:{\rm erg\:s^{-1}}$.

\begin{figure}
	\includegraphics[width=0.48\textwidth]{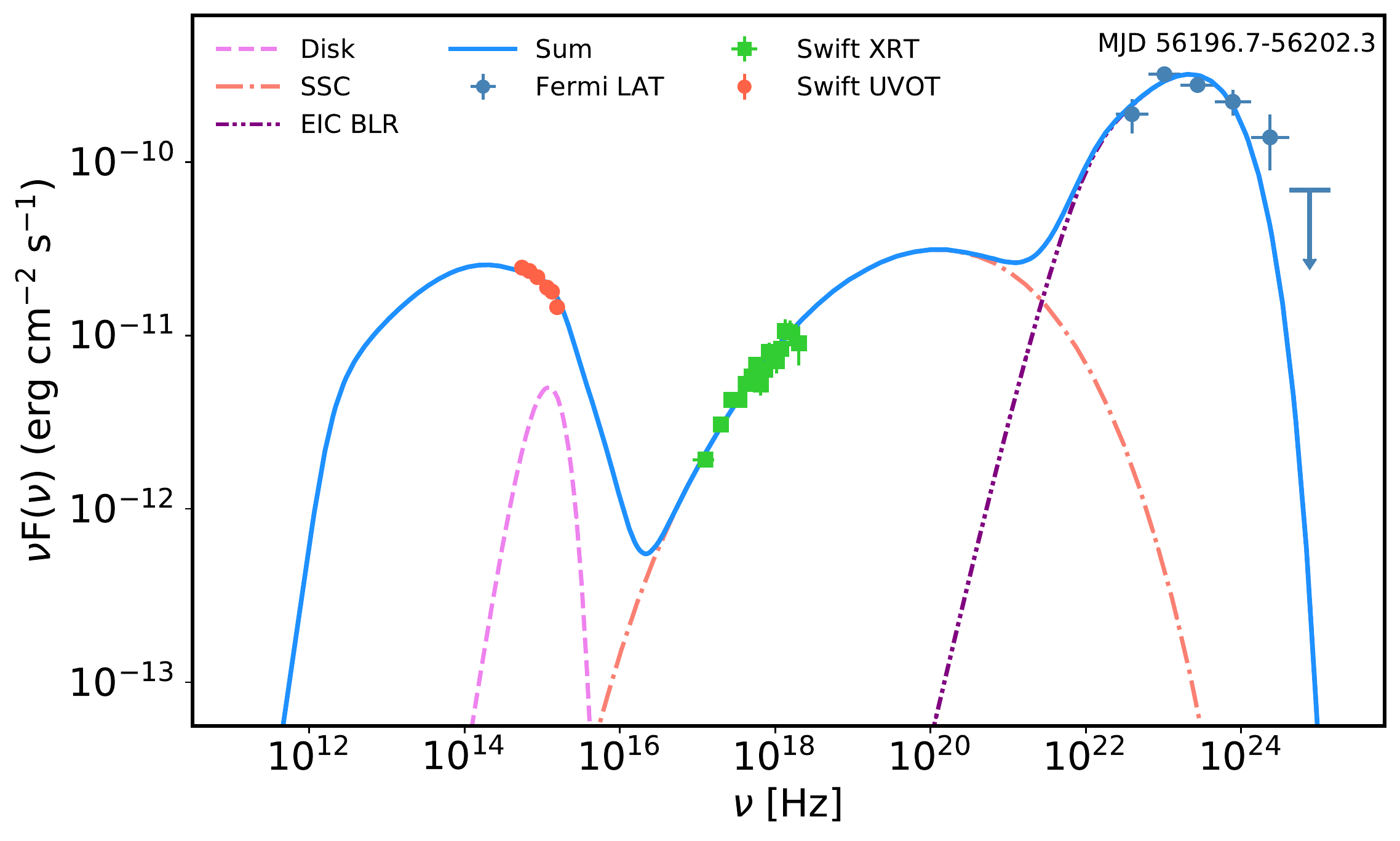}
    \caption{The multiwavelength SED of \source\ during MJD 56196.7-56202.3 constructed with the data from Swift UVOT, XRT and \fermi. The disc, SSC, EIC-BLR and the sum of all components are in dashed pink, dot-dashed orange, dot-dot-dashed purple and solid blue lines, respectively.
    }
    \label{sed}
\end{figure}
To model the broadband SED, a publicly available code, JetSet was used \citep{2006A&A...448..861M, 2009A&A...501..879T, 2011ApJ...739...66T,2020ascl.soft09001T}. JetSet fits the numerical models to observed data and is able to find the optimal values of parameters best describing the data. The multiwavelength SED of \source\ constrained with contemporaneous data observed during MJD 56196.7-56202.3 and modeled with JetSet is shown in Fig. \ref{sed}. The dashed violet line shows the  disc thermal emission approximated as a black body. The power-law index of the emitting electrons is $p=1.61$ while the minimum and cut-off energies are $\gamma_{\rm min}=51.3$ and $\gamma_{\rm cut}=685.6$, respectively. The synchrotron emission of these electrons in the magnetic field of $B=4.43$ G extends up to $10^{16}$ Hz explaining the observed data in the optical/UV bands. Then, the SSC component takes into account the X-ray data (dot-dashed orange curve in Fig. \ref{sed}) dominating only up to $10^{22}$ Hz, failing to explain the \gray\ data. Instead, the inverse Compton upscattering of the BLR photons that have higher mean energy and number density in the jet frame can explain the \gray\ data (dot-dot-dashed purple curve in Fig. \ref{sed}). The modeling allows to estimate the jet parameters such as size of the emission region, $R = 2.03\times10^{15}$ cm and the Doppler factor $\delta=29.8$. The size of the emission region corresponds to the flux variability of the order of $1.3$ hours, consistent with the rapid multi-band variability of \source. 

The modeling of the single snapshot SED shown in Fig. \ref{sed} permits to identify the parameters of the emitting region and the jet for a given period. However, in order to deeply investigate the multiwavelength emission processes in \source\ something beyond the single-epoch SED modeling is required. In \citet{2021MNRAS.504.5074S} and \citet{2022MNRAS.513.4645S} the multiwavelength emission from 3C 454.3 and BL Lac was investigated by modeling as many contemporaneous SEDs as possible constrained during the considered periods. As compared with the single snapshot SED modeling, the advantage of such an approach is that it allows to follow the changes also in the parameters over time, thus get a clue on the evolution of the processes that have lead to the emission in different states (e.g., flares). In addition, such modeling has diagnostic applications, i.e., by fitting many SEDs it is possible to identify periods when the source was characterized with peculiar emission properties that are not possible to explain within the considered model.  

In order to model the SEDs of \source\ in different periods, from the SEDs generated for each Bayesian block there were selected all the periods with sufficient multiwavelength data, i.e., when the optical/UV data at least in two filters is available together with the \gray\ and X-ray data. In Fig. \ref{lightcurve_all} the selected periods are shown in gray. As a result high-quality SEDs in 117 periods were assembled which represent various emitting sates of \source\ including periods when it was in a prolonged flaring state in the \grays. Therefore, this allows to understand the physical processes dominating in the jet of the source in its quiescent and flaring states. All the selected SEDs are modeled within the same one-zone scenario described above. 
\section{Results and Discussion}\label{res}
In this section, the implications of the data analysis are discussed, and the results from the broadband spectral fitting are presented.
In the optical/UV, X-ray and \gray\ bands, \source\ exhibits complex flux changes showing multiple flaring periods. The highest amplitude changes are observed in the HE \gray\ band where the \gray\ luminosity of the source varies from $8.50\times10^{46}\:{\rm erg\:s^{-1}}$ to $7.55\times10^{50}\:{\rm erg\:s^{-1}}$ (assuming a distance of $7.1$ Gpc) which makes \source\ one of the brightest sources in the extragalactic \gray\ sky.

The visual inspection of the multiwavelength light curves in Fig. \ref{lightcurve_all} shows that fluxes in different bands change almost simultaneously. Possible correlation or anticorrelation between the fluxes in different bands shows whether or not the emission is produced by the same population of the particles and related mechanisms. In the case of one-zone leptonic scenario considered here, when the optical/UV photons are from synchrotron emission of the electrons while the emission in the X-ray and \gray\ bands is from the inverse Compton scattering of internal and external photon fields by the electrons in the same emitting region, one expects correlation between the photons at different frequencies as can be seen from  Fig. \ref{lightcurve_all} \citep[e.g.,][]{2019ApJ...880...32L, 2019MNRAS.490..124M, 2020MNRAS.498.5128R}.

\begin{figure*}
	\includegraphics[width=0.48\textwidth]{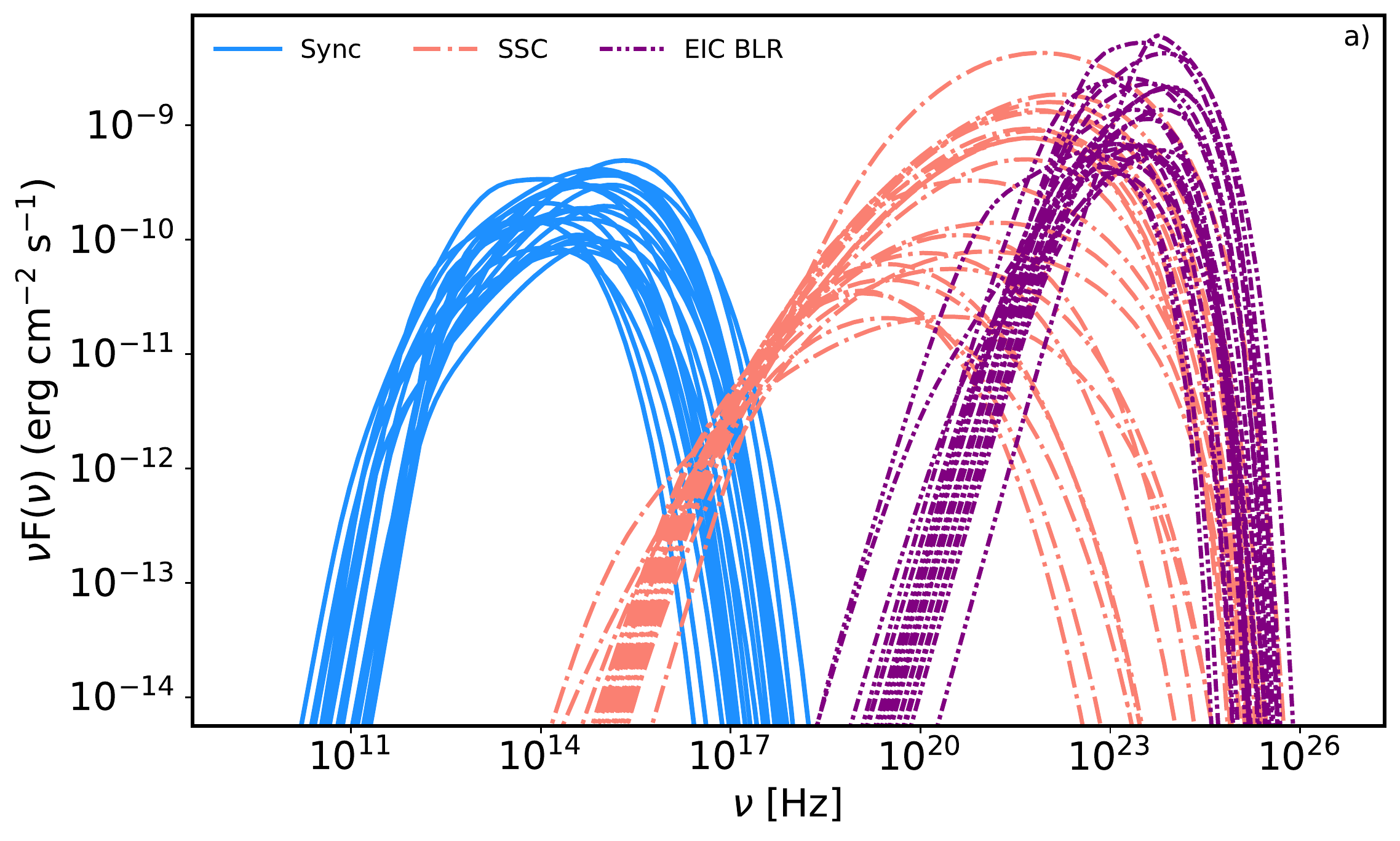}
	\includegraphics[width=0.48\textwidth]{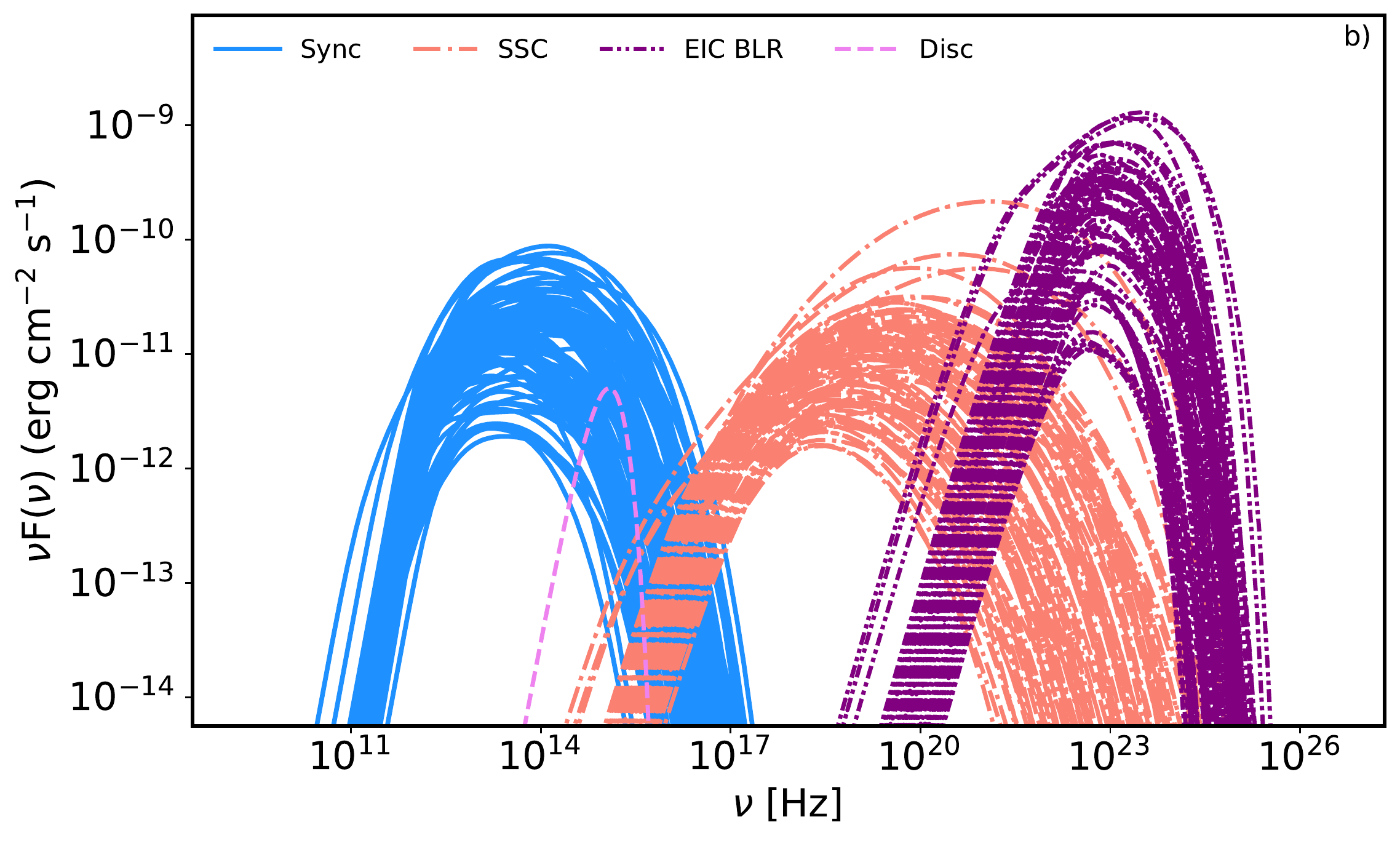}
    \caption{The Multiwavelength SED modeling in different periods. {\it Panel a:} Synchrotron, SSC and EIC components, blue, dot-dashed orange and dot-dot-dashed purple lines, respectively, when the source was in an active state in all the considered bands. {\it Panel b:} The same components in all other periods.}
    \label{sed_model}
\end{figure*}

\subsection{Long-term broadband SED modeling}
The one-zone leptonic model adopted here can adequately reproduce the observed data in almost all the considered periods. The datasets considered here, namely optical/UV, X-ray and \gray\ data, contain relevant information on the source emission in each band, but together they put a constraint on the shape of the emitting particle distribution. Except for the cases when the source is in a very low emission state and the optical/UV emission is (partly) dominated by the thermal emission from the disc, the decaying shape of the optical/UV data directly constrains the HE tail of the synchrotron component which controls the cut-off energy of the emitting electrons ($\gamma_{\rm cut}$). Instead, the X-ray spectrum exhibiting rising shape allows to constrain $p$. Additional constraints on the $\gamma_{\rm cut}$ and $p$ are provided from \gray\ observations: depending on the shape of the \gray\ spectrum, rising, steepening or flat, it defines either the distribution of the particles or their cut-off energy.

The time evolution of the selected SEDs modeling is available here \href{https://youtu.be/0H1IyNN9PSM}{\nolinkurl{youtube.com/0H1IyNN9PSM}}. In Fig. \ref{sed_model} the SED modeling results are shown for each case separating synchrotron (light blue), SSC (dot-dashed orange) and EIC (dot-dot-dashed purple) components. The models are shown by separating the periods when \source\ was in the active states in all the bands (panel a) and in all the other periods (panel b). The low-energy component peaks, as typical for FSRQs, is around $\sim10^{14}$ Hz and is mostly defined by the synchrotron emission of the jet electrons. Although the flux of the synchrotron component varies largely, i.e., in the low state the peak flux can be as low as $\sim10^{-12}\:{\rm erg\:cm^{-2}\:s^{-1}}$ but it can increase up to $\sim10^{-10}\:{\rm erg\:cm^{-2}\:s^{-1}}$ during the flares, the synchrotron peak frequency remains relatively unchanged. However, in several occasions (e.g., between MJD 55228-56190 and MJD 58297-58353) the disc thermal emission with a flux of $\sim 6.86\times10^{-11}\:{\rm erg\:cm^{-2}\:s^{-1}}$ exceeds the synchrotron emission from the jet (violet dashed line in Fig. \ref{sed_model} panel b). As one can see from Fig. \ref{lightcurve_all}, in the mentioned periods the source was in a quiescent state in all the considered bands, so it is natural that bright accretion disc of \source\ overshines the synchrotron component. The relatively constant peak frequency of the synchrotron component limits also the highest energy of the synchrotron photons, and their inverse-Compton scattering steepens in the hard X-ray/soft \gray\ bands, unable to explain the observed \gray\ data (SSC; dot-dashed orange lines in Fig. \ref{sed_model}). Instead, the Compton dominance (the ratio of the high-to-low components luminosity) and the \gray\ spectra are naturally explained by inverse Compton scattering of BLR photons (EIC; dot-dot-dashed purple lines in Fig. \ref{sed_model}).

\begin{figure}
	\includegraphics[width=0.48\textwidth]{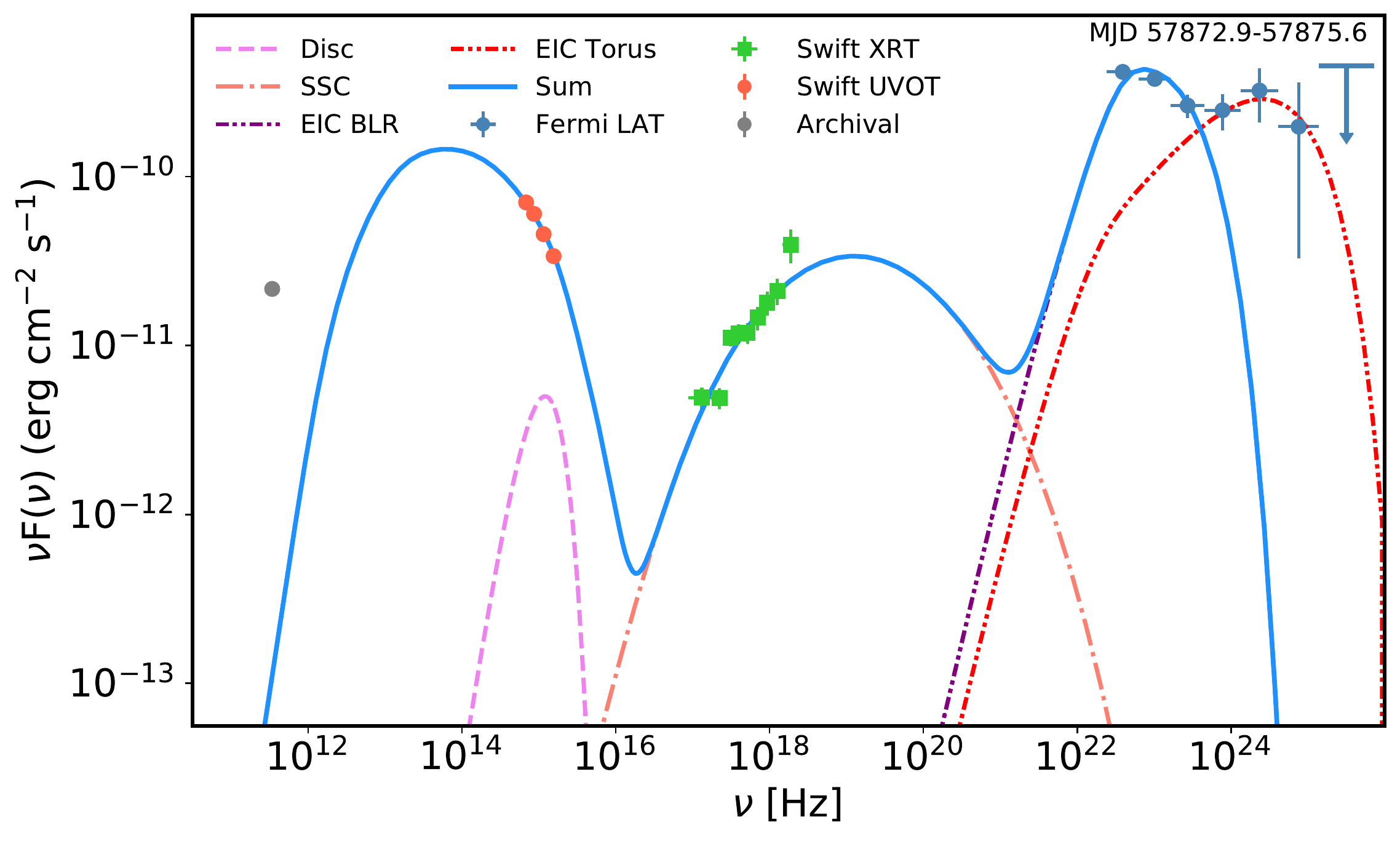}
    \caption{The multiwavelength SED of \source\ during MJD 57872.9-57875.6 when the \gray\ spectrum was flat, extending up to 58 GeV. The same color code as in Fig. \ref{sed} is adopted. The light red dot-dot-dashed line shows EIC torus component when the second emitting region is outside BLR.}
    \label{sed270}
    \end{figure}

The models shown in panels a) and b) of Fig. \ref{sed_model} demonstrate different behaviour of the \source\ emission in active and other states. The brightening of the source substantially modifies different components affecting their flux and spectrum. For example, when modeling the SED in the bright X-ray state characterized by a hard X-ray photon index, the intensity of the SSC component increases and its spectrum hardens extending the peak of this component to higher energies (panel a) Fig. \ref{sed_model}). However, for a harder photon index (hence a lower $p$), $\gamma_{\rm cut}$ should be lower not to violate the optical/UV data. So, even in those bright and hard X-ray states the SSC component has a decreasing shape in the GeV band, and again the \fermi\ observed data are interpreted as inverse-Compton up-scattering of BLR emitted photons. Similarly, the spectral variability in the MeV/GeV band affects the EIC component. As an example, the SED of \source\ during MJD 57872.9-57875.6 is shown in Fig. \ref{sed270}. During this period, the MeV/GeV spectrum is characterized by a nearly flat spectrum extending up to 58 GeV. The modeling shows that the distribution of the emitting electrons is described by $p=1.87$ power-law index and the cut-off energy of $\gamma_{\rm cut}=306.4$. So, the inverse Compton scattering of BLR photons can reach only 2 GeV unable to explain the observed data in tens of GeV. The limit imposed by the emitting electron distribution prohibits the interpretation of GeV data within one-zone scenarios; the observed data can be account for only when the photons with higher mean energy are inverse Compton up-scattered on the same electrons. In the emitting region, except for BLR, the electron can interact with disc photon or the photons emitted from the dusty torus. The inverse Compton scattering of the disc photons will produce a peak comparable to that shown in Fig. \ref{sed270} whereas in the case of the dusty torus photons with a lower mean energy will produce a peak at lower frequencies. The emission in the $>2$ GeV band is most likely produced from the second emission region containing more energetic electrons. As an example, in Fig. \ref{sed270} the GeV data are modeled as emission from the second region which is assumed to be be outside the BLR. As the data are not sufficient to constrain the parameters, it is assumed that this region \textit{i)} has the same Doppler boosting factor ($\delta=29.4$) as the one inside the BLR (constrained from the fit), \textit{ii)} is characterized by a significantly lower magnetic field ($0.2$ G as compared to $12.3$ G estimated for the other region) not to overproduce the X-ray data which are from the region within the BLR and \textit{iii)} contains more energetic electrons with $p=1.80$ and $\gamma_{\rm cut}=1.10\times10^{4}$. As the emitting region is outside the BLR, the dominant photon field is IR photons from the dusty torus; the inverse Compton up-scattering of these photons is shown with a light red dashed line in Fig. \ref{sed270} which extends up to GeV bands and accounts for the observed data. In principle, the second emission region can be a local structure in the jet where the particles are re-accelerated (e.g., a local reconnection outflow in the “jet in a jet” scenario \citep{2009MNRAS.395L..29G, 2010MNRAS.402.1649G}) or there occurs an injection of fresh electrons. The modeling presented above is to show that the observed data in some cases (e.g., when the \gray\ spectrum is flat and extends to tens of GeV, two among the selected SEDs) cannot be reproduced in one-zone scenarios, so that more complex (e.g., two-zone) scenarios are required.
\subsection{Energy distribution of the emitting electrons}
\begin{figure*}
	\includegraphics[width=0.48\textwidth]{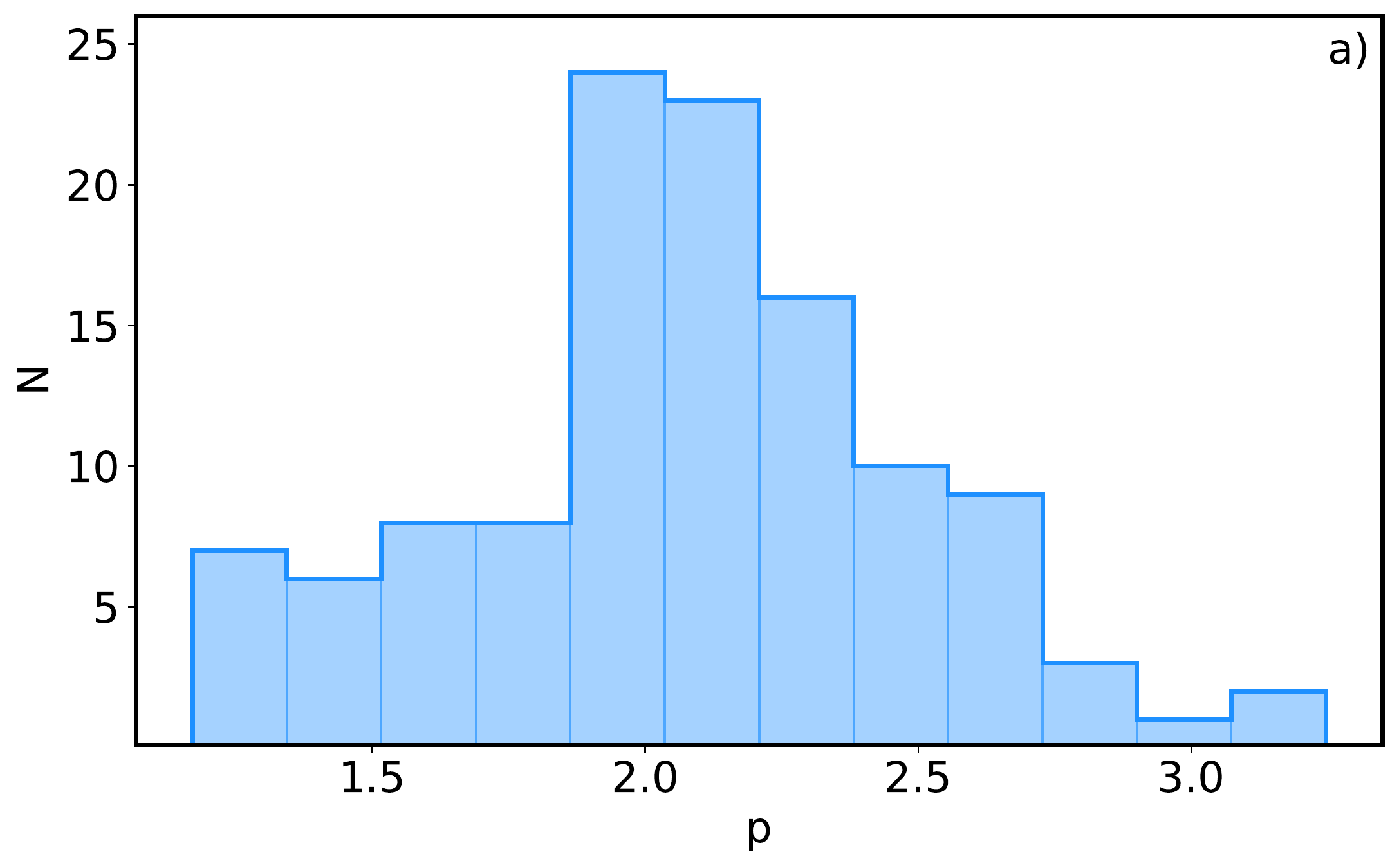}
	\includegraphics[width=0.48\textwidth]{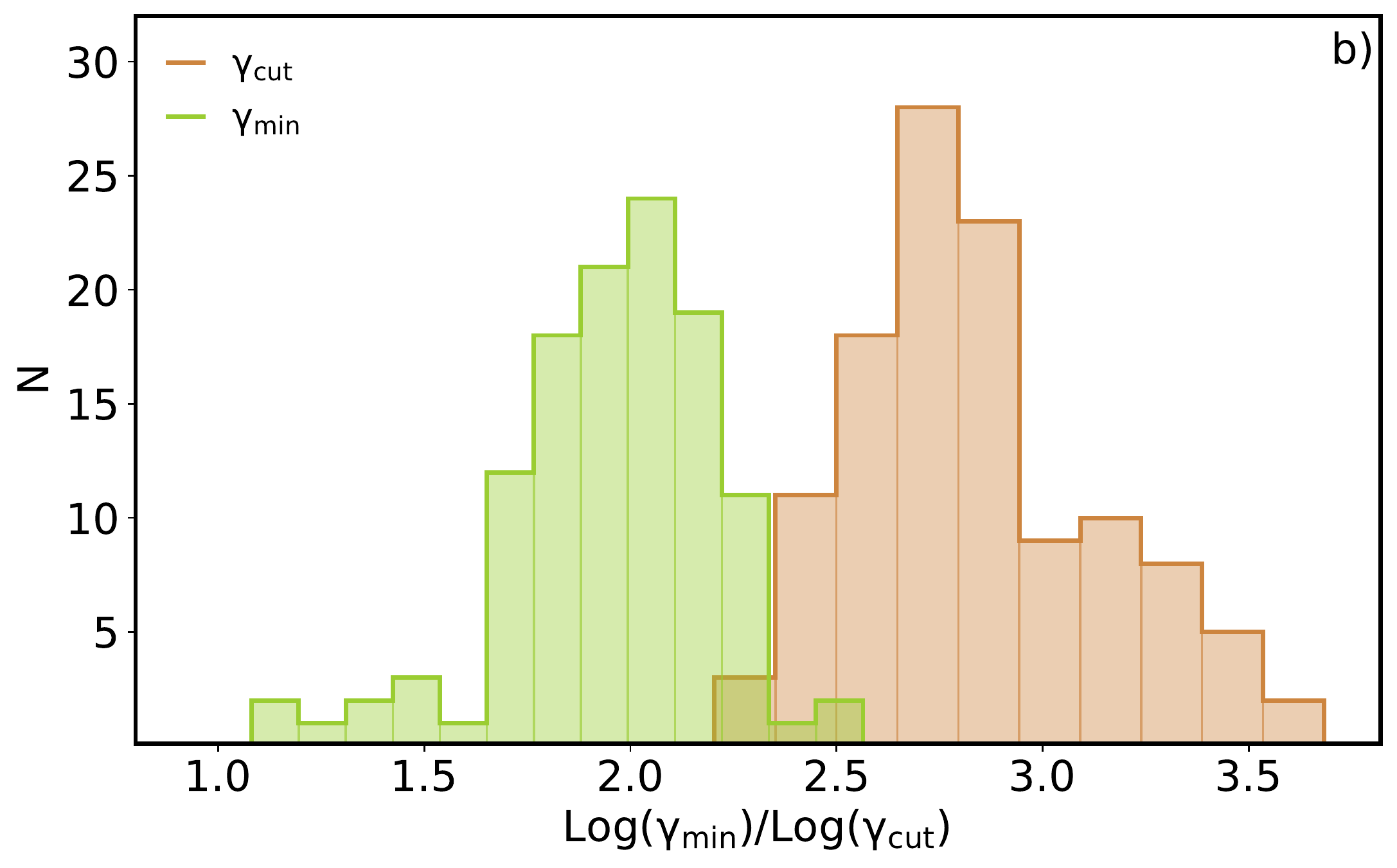}\\
	\includegraphics[width=0.48\textwidth]{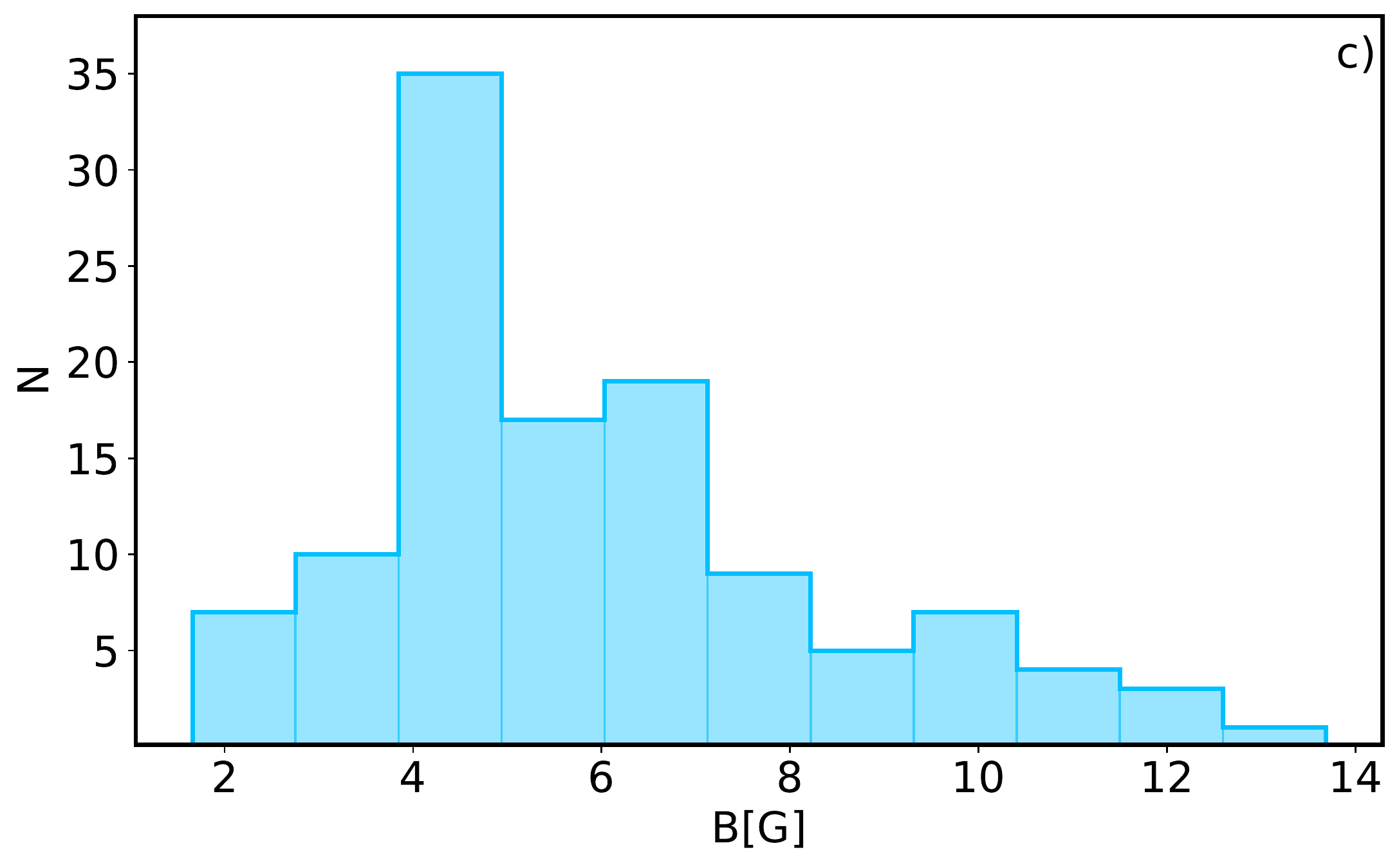}
	\includegraphics[width=0.48\textwidth]{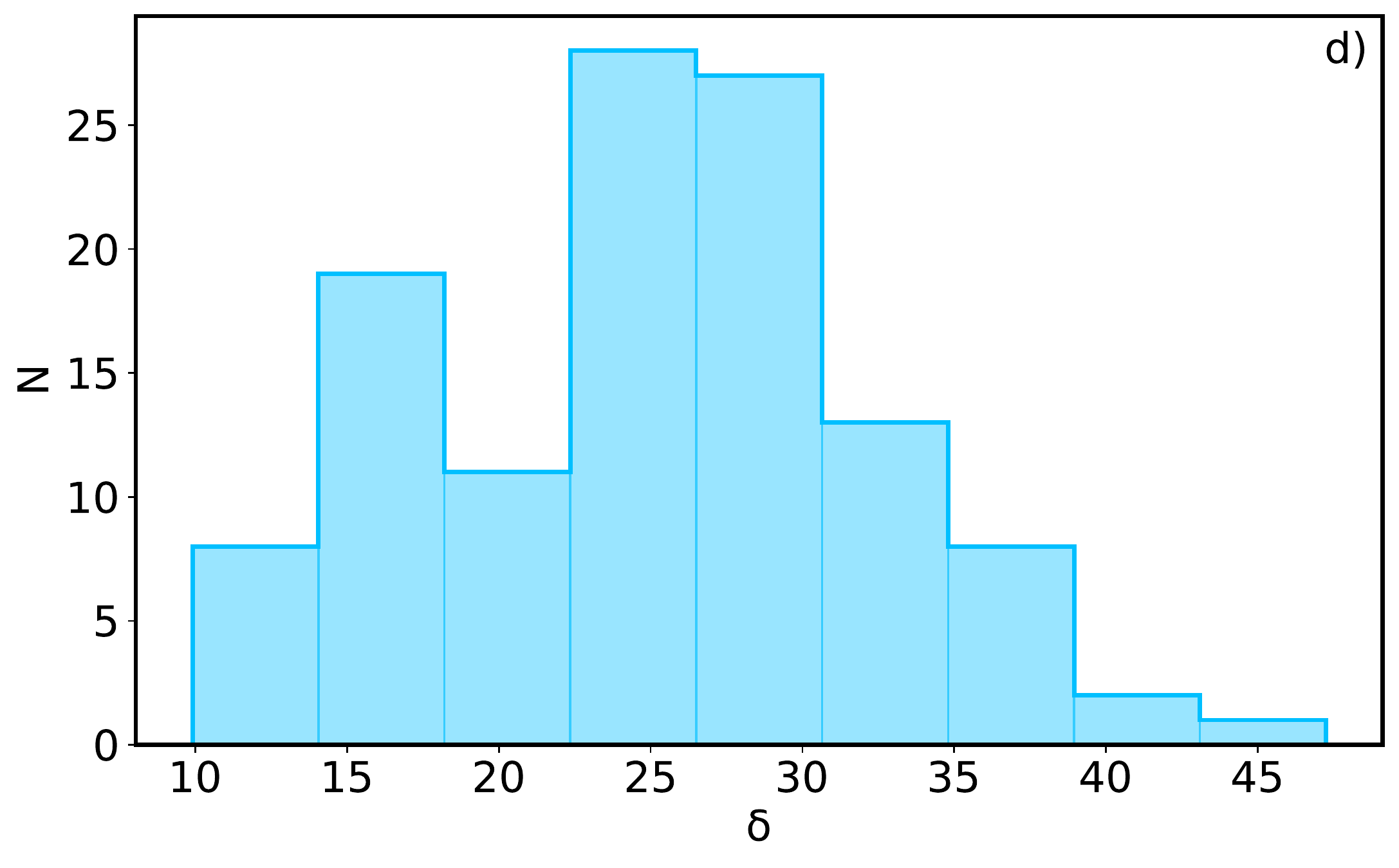}\\
	\includegraphics[width=0.48\textwidth]{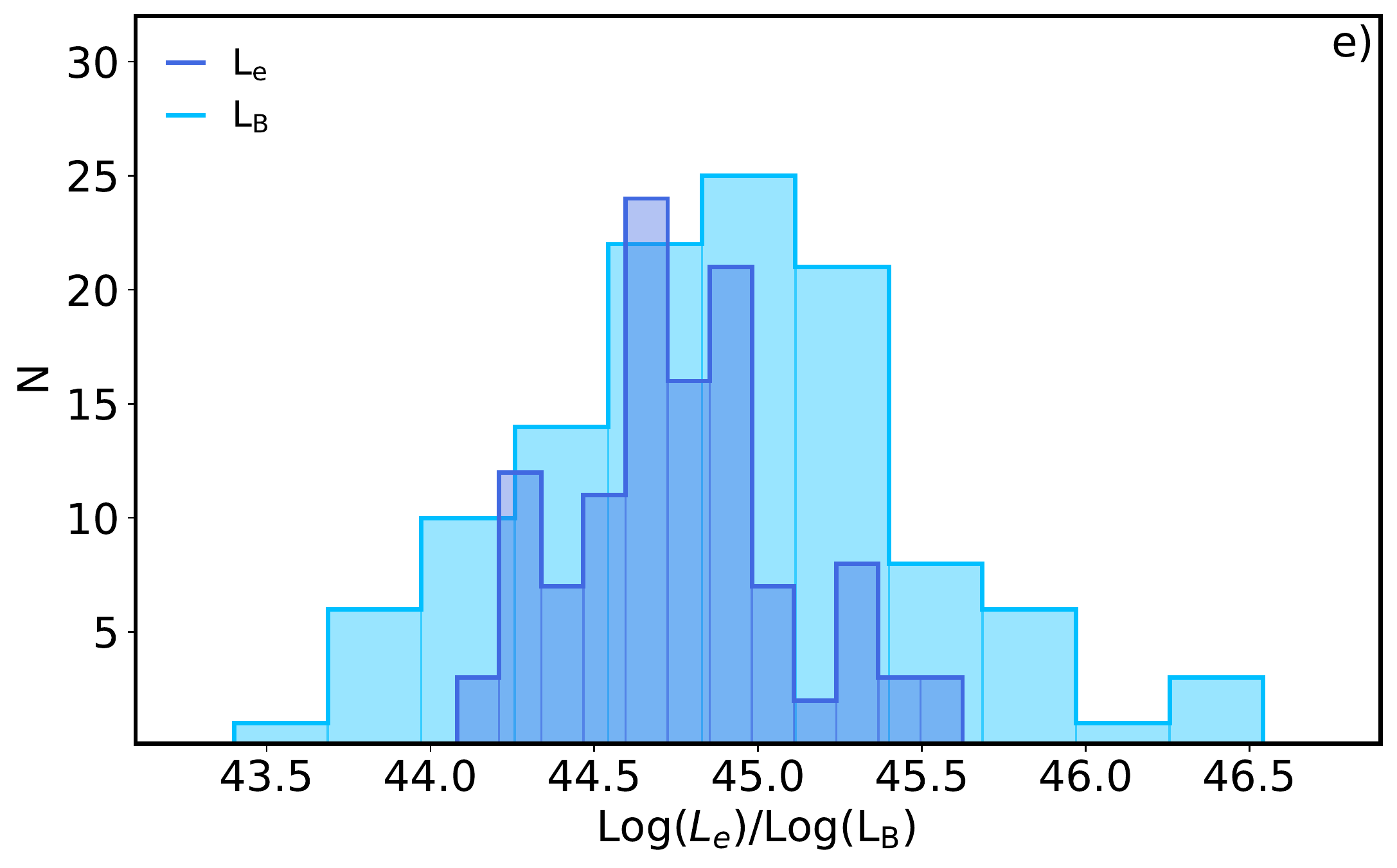}
    \caption{The distribution of the parameters obtained from the fitting of all data-sets composed with simultaneous data. \textit{a)} The distribution of the emitting electron power-law index, \textit{b)} the distribution of $\gamma_{\rm min}$ (green) and $\gamma_{\rm min}$ (orange),  \textit{c)} magnetic field distribution, \textit{d)} Doppler factor distribution and \textit{e)} the distribution of $L_{\rm e}$ (dark blue) and $L_{\rm B}$ (light blue).}
    \label{sed_histo}
\end{figure*}
The modeling of 117 high-quality SEDs of \source\ with diverse features allows to investigate the properties of the jet and emitting particles over time. In Fig. \ref{sed_histo} the distribution of $p$, $\gamma_{\rm min}$, $\gamma_{\rm cut}$, $B$, $L_{\rm e}$ and $L_{\rm B}$ obtained from the modeling are shown. The wide distribution of the considered parameters once more shows the complex changes having taken place in the jet of \source. The power-law index of the emitting electron distribution varies between $p=1.17-3.25$ with a mean of $p_{\rm mean}=2.08$ (Fig. \ref{sed_histo} panel a). This power-law index constrained by the X-ray and \gray\ data varies following the spectral changes in the X-ray and \gray\ bands; when a steep falling spectrum is observed in the \gray\ band, the emitting electron should also have a steep spectrum, while $p<2.0$ are expected in bright active states that are characterized by a hard photon index. The distribution of $\gamma_{\rm min}$ and $\gamma_{\rm cut}$ is shown in Fig. \ref{sed_histo} panel b). Both parameters have a narrow distribution peaking around $\gamma_{\rm min, mean}=104.6$ and $\gamma_{\rm cut, mean}=905.1$, respectively. The narrow distribution of $\gamma_{\rm cut}$ (between $(1.60-48.16)\times10^{2}$) is probably due to stability of $\nu_{\rm peak}$ but in general it depends also on $p$. The magnetic field estimated in different periods (Fig. \ref{sed_histo} panel c), varies from $B=1.66$ G to $B=13.69$ G with a mean of $B_{\rm mean}=5.96$ G. For example, the highest magnetic field of $B=13.69$ G was estimated from fitting the SED observed between MJD 57754-57756 when the source was in an elevated optical/UV emission state. So, the increase of the synchrotron component leads to an increase in $B$: large magnitude change of the synchrotron component can be seen from Fig. \ref{sed_model}.

The distribution of $\delta$ in different periods is shown in Fig. \ref{sed_histo} panel d). The high values of $\delta$ are mostly estimated during the flares in the \gray\ band, for example, the highest value of $\delta=47.2$ was observed on MJD 57743.2 when the source was in a \gray\ active state.
It should be noted that sometimes high values of $\delta$ have already been estimated for \fermi\ detected blazars \citep[e.g., see][]{2020ApJ...897...10Z} and are usually used to model the bright blazar flares observed in HE or VHE \gray\ bands \citep[e.g., see][]{2019A&A...627A.159H, 2019ApJ...884..116L}. When $\delta$ increases, a lower electron density is required to produce the same level of synchrotron radiation, so the synchrotron photon density and the SSC component decrease but the external photon energy density in the jet frame becomes larger leading to the increase of the EIC component. For this reason, the enhancement in the \gray\ band results in higher $\delta$.

The parameters distribution presented in Fig. \ref{sed_histo} does not differ from that usually estimated for \source\ in different periods. For example, in \citet{2018ApJ...863..114G} by considering different locations of the emission region it is found that SSC and EIC of BLR photons can explain the broadband SED in the low state when $p=2.51\pm0.11$, $\gamma_{\rm cut}=1311.1\pm195$, $B=5.40\pm0.13$ and $\delta=10$. Whereas in the active state, the data can be explained when these parameters are: $p=1.81\pm0.09$, $\gamma_{\rm cut}=724.1\pm78$, $B=8.24\pm0.18$ and $\delta=30$ \citep{2020A&A...635A..25S}. Or in \citet{2018ApJ...866...16P} by assuming a log-parabolic electrons injection spectrum, it is shown that in a  pre-flare state the SED of \source\ can be modeled when the injection index of the electrons is $1.9$ and the curvature is $0.08$ but in the flaring states the index becomes $1.7$ with a curvature of $0.02$. The magnetic field is estimated to be around $4$ G. Moreover, it should be noted that there are other models which explain the flaring activity of \source, e.g., those considering the ablation of a gas cloud penetrating the relativistic jet and computing the expected multiwavelength emission from the leptonic and hadronic interactions, see \citet{2017ApJ...851...72Z} and \citet{2019ApJ...871...19Z}. Also, the parameters estimated within these models are not significantly different from those presented in Fig. \ref{sed_histo}.
\subsection{Formation of electron spectrum}
\begin{figure}
	\includegraphics[width=0.48\textwidth]{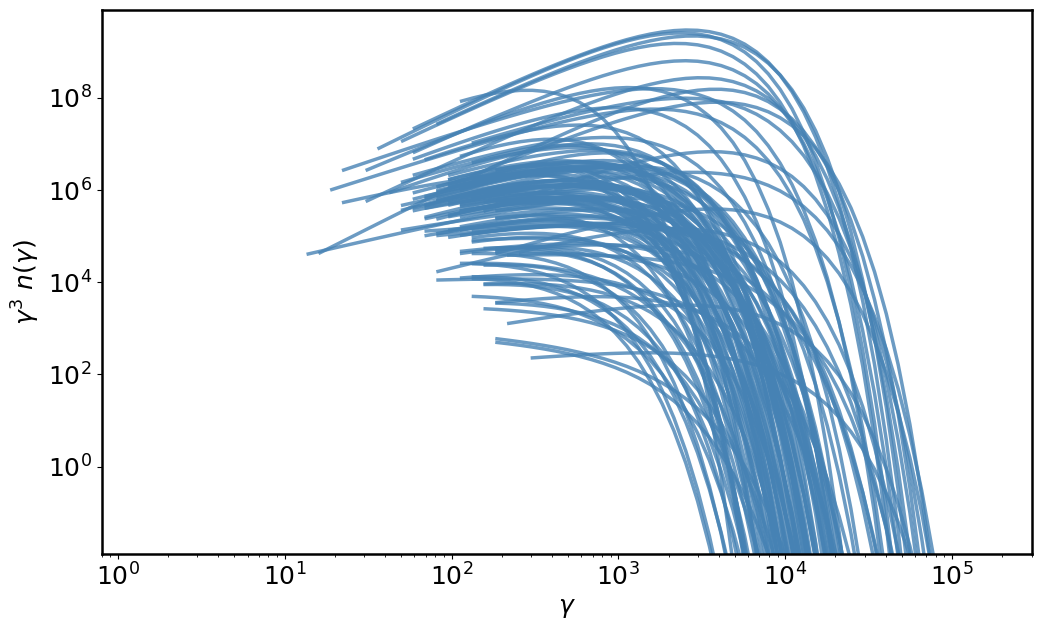}
	\includegraphics[width=0.48\textwidth]{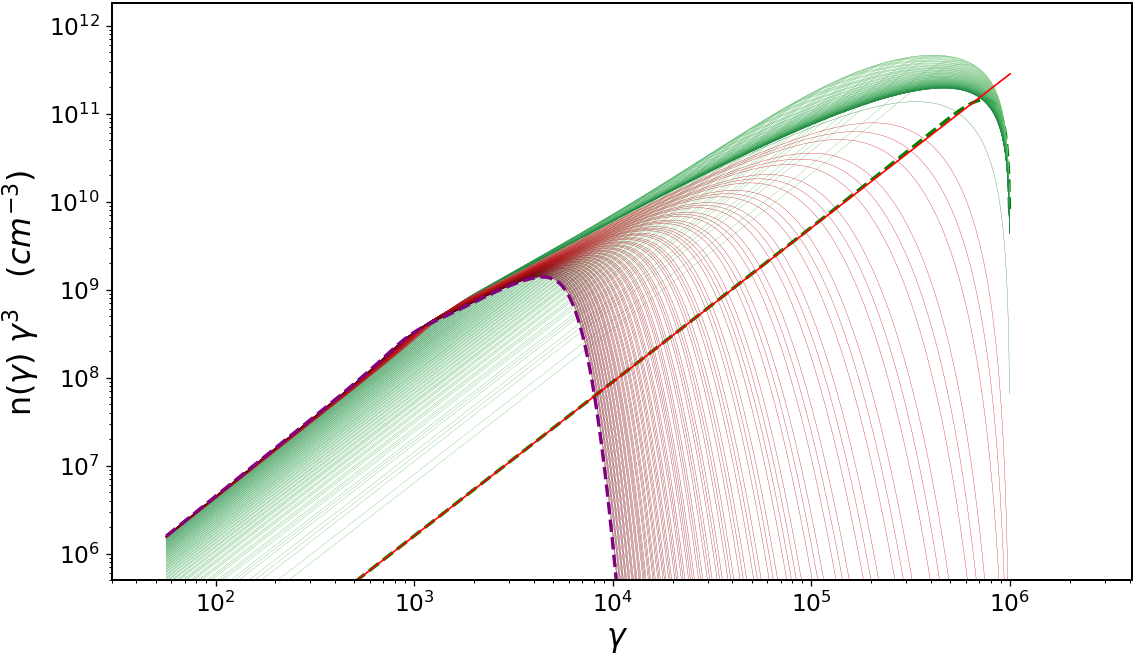}
    \caption{{\textit Upper panel:} Electron energy distributions obtained from modeling of SEDs. {\textit Lower panel:} The evolution of the energy spectrum of the electrons injected in the emitting region. Red and green lines show the electron spectrum in different steps and the final spectrum is shown in purple.}
    \label{elect_spec}
\end{figure}
The electron spectrum given in Eq. \ref{eldist} is an ad-hoc assumption of the distribution of particle injected in the emitting region. This approach, however, ignores the formation of the particle spectrum which is governed by different cooling processes and gains through particle energization mechanisms. From the theoretical point of view, the mechanisms usually considered for the particle acceleration are shock acceleration \citep[e.g.,][]{2008ApJ...682L...5S, 2012ApJ...745...63S, 2017MNRAS.464.4875B} or magnetic reconnection \citep[e.g.,][]{2001ApJ...562L..63Z,2014ApJ...783L..21S,2015ApJ...806..167G}. However, all the considered mechanisms to some degree face difficulties to explain all the constraints imposed from the multiwavelength SED modelings. Here we do not attempt to discuss the exact mechanisms that have led to the particle acceleration and injection in the emitting region but instead we investigate whether or not the distribution of the electron spectrum necessary to model the broadband SEDs of \source\ can be formed under the physical conditions considered above. A more straightforward approach to gain much information on the particle acceleration and cooling mechanisms would be self-consistent consideration of particle spectrum from acceleration to cooling and comparing its radiative signature with the multiwavelength data. This will be studied in a future paper. 

Fig. \ref{elect_spec} upper panel shows the distributions of the electrons estimated from the modeling of selected SEDs. This clearly demonstrates different properties of the emitting particles and their evolution in time. In particular, the spectrum of the electrons sometimes is hard ($p<2.0$) and extends above $\gamma_{\rm e}>10^{3}$ however steep and narrow distributions were also obtained. The power-law index of the electron distribution directly points to the acceleration mechanisms which is unknown while $\gamma_{\rm cut}$ is due to the interplay of acceleration and cooling processes. In order to calculate the temporal evolution of the electron spectrum, an integro-differential equation that takes into account the injection, cooling (considering all the radiative fields) and escape of the particles should be solved \citep{1962SvA.....6..317K}. This is done using {\it JetTimeEvol} class of the JetSet. This class numerically solves the kinetic equation and allows to evolve the particle distribution under any cooling process. 

In the electron distribution the limiting factors constraining $\gamma_{\rm cut}$ are the efficiencies of the acceleration process (namely the acceleration/injection time $t_{\rm inj}$) and the physical size of the accelerator. In other words, the electrons will not be accelerated beyond the energies when the radiative cooling time ($3/4\:c\:\sigma_{\rm T}\:\frac{U_{\rm tot}}{m_e\:c^2}\:\gamma^2$, where $U_{\rm tot}$ is the sum of magnetic and photon fields) is shorter than the acceleration time. In the one-zone scenario considered here when the emission region is within the BLR, the electrons are cooled through interaction with the magnetic and photon fields, so $U_{\rm tot}$ is synchrotron plus photon energy density, i.e. $U_{\rm tot}=U_{\rm B}+U_{\rm SSC}+U_{\rm EIC}$.
In order to discuss the evolution of the particle distribution in time, we assume that power-law distributed electrons with $p=1.25$ are injected into the emitting region where the magnetic field is $4.1$ G and $\delta=24.2$. These are chosen to be similar to the parameters estimated from the SED modeling observed during MJD 57715.6-57716.8 (see the SED modeling animation) when the source was in an active emission state. In this case, the synchrotron cooling time for the electrons with energy of $\gamma_{\rm e}=10^4$ is $t_{\rm syn, cool}=4\times10^3$ s. The evolution of the energy spectrum of electrons with a luminosity of $L_{\rm e}=1.74\times10^{45}\:{\rm erg\:s^{-1}}$ injected into the emitting region with a radius of $2.38\times10^{15}$ cm and without escape is shown in Fig. \ref{elect_spec} lower panel. The red dashed line corresponds to the initial injection spectrum of the electrons. As the cooling time is inverse proportional to the energy of the electrons, initially only the highest energy electrons are cooled down, forming a turnover (cut-off) in the spectrum. In time, this cut-off energy gradually moves to lower energies and when the injection time is $\simeq2.6\times10^3$ s the cut-off energy will be around $1.5\times10^3$ close to the value estimated from SED modeling. In time, however, this cut-off energy will move to lower ranges.  

\begin{figure}
	\includegraphics[width=0.48\textwidth]{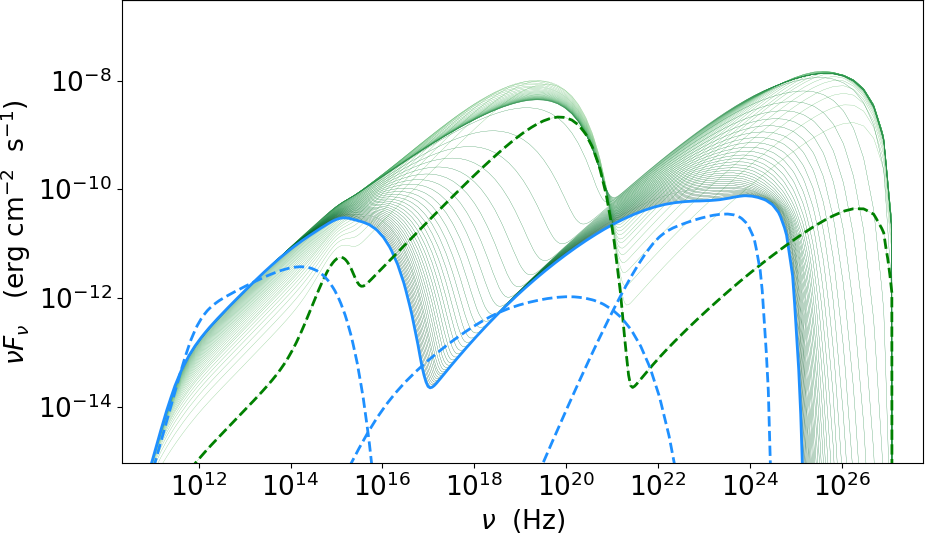}
    \caption{The SED evolution in time after the injection of the power-law electrons. The green dashed line shows the initial SED, the green solid lines show the SED in different steps while the final SED is in solid blue. The dashed blue line shows the SED for longer evolution of the system.}
    \label{sed_inj}
\end{figure}

When the injected electrons start to cool, their radiative signature changes in time. The SEDs corresponding to electron spectra given in Fig. \ref{elect_spec} lower panel are shown in Fig. \ref{sed_inj}. The SED of initially injected electrons (the sum of synchrotron, SSC and EIC components) is shown with green dashed line. This spectrum modifies in time when the injected electrons start to cool; the green solid lines show the evolution of the sum of all component in time which shows that the synchrotron and inverse Compton peaks move to lower frequencies. By cooling, the highest energy electrons are transferred to lower energies, so the number of low-energy (i.e., not cooled) electrons changes and their synchrotron emission increases at lower frequencies (e.g., around $10^{12}$ Hz). Similarly, the SSC component increases in the X-ray band, while EIC dominates in the HE \gray\ band. The blue line in Fig. \ref{sed_inj} is the final SED produced from the electron population with a spectrum shown by a purple line in Fig. \ref{elect_spec} lower panel. It matches with that obtained from the modeling of SED observed on MJD 57715.6-57716.8 when using electron distribution given by Eq. \ref{eldist}. For later periods, the resulting spectrum decreases in intensity and moves to lower frequencies which is shown as a blue dashed line in Fig. \ref{sed_inj}. The resulting spectrum is more characteristic to source emission when it is in quiescent state. Therefore, the electron spectra obtained from the fitting of SEDs can be naturally formed in time.
\subsection{Jet power}
The modeling allows also to estimate the jet power carried by electrons ($L_{\rm e}$) and magnetic field ($L_{\rm B}$). The distribution of the luminosities computed as $L_{e}=\pi c R_b^2 \Gamma^2 U_{e}$ and $L_{B}=\pi c R_b^2 \Gamma^2 U_{B}$ is shown in Fig. \ref{sed_histo} panel e). The mean of $L_{e}$ and $L_{B}$ is at $7.81\times10^{44}\:{\rm erg\:s^{-1}}$ and $2.07\times10^{45}\:{\rm erg\:s^{-1}}$, respectively. The distribution of $L_{B}$ in the range $2.51\times10^{43}-3.48\times10^{46}\:{\rm erg\:s^{-1}}$ is broader than that of $L_{e}$ between $1.20\times10^{44}-4.21\times10^{45}\:{\rm erg\:s^{-1}}$. The large variations of $L_{B}$ are mostly due to the high-amplitude changes of the synchrotron component in the SED of \source. Instead, the high-amplitude increase of the \gray\ flux interpreted as EIC of BLR photons which would affect the electron content in the jet is compensated by increasing $\delta$. The distribution of $L_{e}$ and $L_{B}$ in Fig. \ref{sed_histo} panel e) shows that in some periods $L_{e}/L_{B}<1$, i.e., the jet is magnetically dominated. Such a trend is observed when the synchrotron component (defined by optical/UV data) exceeds the SSC component (defined by X-ray data). 

The estimated parameters allow also to asses the total kinetic energy of the jet, namely, assuming a proton-to-electron comoving number density ratio of $N_{p}/N_{e}\simeq0.1$, the total kinetic luminosity defined as $L_{\rm kin}=L_{e}+L_{B}+L_{\rm p, cold}$ varies from $4.64\times10^{44}\:{\rm erg\: s^{-1}}$ to $3.71\times10^{46}\:{\rm erg \:s^{-1}}$. Similarly, when $N_{p}/N_{e}\simeq0.01$ and $N_{p}/N_{e}\simeq0.5$, $L_{\rm kin}$ varies from $2.72\times10^{44}\:{\rm erg\: s^{-1}}$ to $3.66\times10^{46}\:{\rm erg \:s^{-1}}$ and from $1.08\times10^{45}\:{\rm erg\: s^{-1}}$ to $3.93\times10^{46}\:{\rm erg \:s^{-1}}$, respectively. The central black hole mass in \source\ is estimated to be $8.5\times10^8\:{\rm M_{\rm BH}}$ \citep{2014Natur.510..126Z}, so the Eddington luminosity is $\simeq1.1\times10^{47}\:{\rm erg\:s^{-1}}$. Therefore, the kinetic power of the jet estimated in various periods is lower than the Eddington luminosity.

\section{Conclusion}\label{conc}

In this paper we studied the physical processes taking place in the jet of \source\ using the results from long-term (fourteen-year-long) multiwavelength observations. We systematically studied the features of the source emission in optical/UV, X-ray and \gray\ bands. Generating the \gray\ light curve with the help of an adaptive binning method, the high-amplitude, multiple flaring and complex variability of the source is investigated. 

The broadband emission from \source\ was investigated by modeling 117 high-quality SEDs assembled during the considered period. This new comprehensive approach allowed to compare and contrast jet and emitting particle properties in different states of the source emission as well as follow the dynamical changes of the physical processes governing in the jet. The one-zone model, when the low energy emission is due to synchrotron radiation of electrons while HE is due to inverse Compton scattering of both synchrotron and BLR reprocessed photons, adequately explains the source emissions in different periods, except the cases when the \gray\ spectrum is flat, extending to tens of GeV (2 out of 117 periods). It is found that during the flaring periods the spectrum of the emitting electrons has a harder distribution and they are effectively accelerated up to $\gamma_{\rm cut}=(1-4)\times10^{3}$ as opposed to the other periods when the electrons have narrow energy distributions. By modeling also the jet kinetic power was assessed showing that it always remained below the Eddington power.
%
\section*{Acknowledgements}
We acknowledge the use of data, analysis tools and services from the Open Universe platform, the ASI Space Science Data Center (SSDC), the Astrophysics Science Archive Research Center (HEASARC), the Fermi Science Tools, the All-Sky Automated Survey for Supernovae (ASAS-SN), the Astrophysics Data System (ADS), and the National Extra-galactic Database (NED).

This work was supported by the Science Committee of RA, in the frames of the research project No 20TTCG-1C015. Also, this work was made in part by a research grant from the Armenian National Science and Education Fund (ANSEF) based in New York, USA.

This work used resources from the ASNET cloud.
\section*{Data availability}
All data used in this paper is public and is available from the \textit{Swift}, Fermi and \textit{NuSTAR} archives and from the Open Universe tools.The V- and R- band data from Steward Observatory, V-band data from ASAS-SN and the V-band data from CSS used in this article are publicly available at \url{http://james.as.arizona.edu/~psmith/Fermi/}, \url{https://asas-sn.osu.edu/} and \url{http://nunuku.caltech.edu/cgi-bin/getcssconedbid_release2.cgi}, respectively. The \fermi, \textit{Swift} XRT and UVOT data analyzed in this paper as well as multiwavelength SEDs in different periods can be shared on reasonable request to the corresponding author.



\bibliographystyle{mnras}
\bibliography{biblio} 








\bsp	
\label{lastpage}
\end{document}